\documentclass[aps,prl,reprint,superscriptaddress]{revtex4-1}
\usepackage{amsmath}
\usepackage{amsfonts}
\usepackage{graphicx}
\usepackage{bm}
\usepackage{braket}
\usepackage{url}
\usepackage{parskip}
\usepackage{orcidlink}

\begin{document}
\title{Classical and Quantum Analysis of Light Transmission Through Polarizing Filters}
\author{D. A. Assunção\orcidlink{0000-0002-6235-0159}}
\affiliation{Institute of Physics, Institute of Exact Sciences, Federal University of Alfenas}
\author{S. B. Soltau\orcidlink{0000-0002-7211-2533}}
\affiliation{Institute of Physics, Institute of Exact Sciences, Federal University of Alfenas}
\date{\today}

\begin{abstract}\noindent
We analyze the behavior of light as it passes through systems composed of polarizing filters at different angular orientations. The analysis is initially conducted in the context of classical optics, using Malus's Law to calculate the transmitted intensity for the cases of two perpendicular filters ($0^\circ$ and $90^\circ$) and three filters ($0^\circ$, $45^\circ$, and $90^\circ$). We then apply quantum mechanics to describe the phenomenon using state notation and the probabilistic interpretation of the projection of the polarization states of photons.
\end{abstract}

\maketitle

\section{Introduction}

The behavior of light passing through polarizing filters has been a fundamental topic in physics. The description of the physical phenomenon of polarization contrasts classical and quantum interpretations of the phenomenon. The classical version, particularly through Malus's law, describes the intensity of polarized light as it passes through a polarizer. In contrast, quantum mechanics offers an interpretation based on superposition and the projection of polarization states of photons~\cite{Wodkiewicz1985}, revealing unique insights into polarization manipulation.

In the classical electrodynamics approach, polarization is defined as the dipole moment per unit volume, applicable to various charge distributions~\cite{Tsukerman2021}. Malus's law is derived from this framework, emphasizing the linear relationship between the polarization angle and the transmitted intensity. The quantum mechanics approach introduces complexities, such as entangled photons, where Malus's law can be generalized to explain correlations beyond classical predictions~\cite{Goncharov2023}. Quantum interpretations also address phenomena such as frequency shifts in X-ray radiation, which classical models overlook~\cite{Volobuev2013}.

Recent studies propose innovative methods, such as metasurfaces, to enhance polarization encoding and multiplexing, expanding the applications of Malus's law in both classical and quantum optics~\cite{Qiu2024, Liangui2020}. Such advancements enable sophisticated control over the properties of light, facilitating applications in cryptography and imaging generation. While the classical view simplifies the concepts of polarization, the quantum perspective reveals deeper complexities and potential applications, suggesting a rich interaction between the two prescriptions.

This article compares these two approaches for two scenarios of polarizing filters: (\textit{i}) two perpendicular filters ($0^\circ$ and $90^\circ$); and (\textit{ii}) three filters, with one intermediate at $45^\circ$. Mathematical descriptions compatible with both classical and quantum versions are adopted.

\section{Classical Analysis}

\subsection{Case of Two Perpendicular Filters}

Consider unpolarized light with initial intensity $I_{0}$ incident on a system of two polarizing filters, $A$ and $C$, oriented at $0^\circ$ and $90^\circ$, respectively. When unpolarized light passes through the first polarizing filter $A$ ($0^\circ$), the intensity of the transmitted light is reduced by half as shown in Eq.~\eqref{eq:eq01}:
\begin{equation}
\label{eq:eq01}
I_{A} = \frac{I_{0}}{2}.
\end{equation}
After passing through filter $A$, the light is polarized in the $0^\circ$ direction. When passing through filter $C$, which is oriented at $90^\circ$, the intensity of the transmitted light is governed by Malus's Law, as indicated in Eq.~\eqref{eq:eq02}:
\begin{equation}
\label{eq:eq02}
I_{C} = I_{A} \cos^{2}(90^\circ).
\end{equation}
Since $\cos(90^\circ) = 0$, we have the following expression for the transmitted intensity after filter $C$ (Eq.~\eqref{eq:eq03}):
\begin{equation}
\label{eq:eq03}
I_{C} = I_{A} \times 0 = 0.
\end{equation}
Therefore, the intensity of the light after filter $C$ is zero. The light polarized at $0^\circ$ has no components in the $90^\circ$ direction, resulting in its complete absorption by filter $C$.

\subsection{Case with Intermediate Filter at $45^\circ$}

Now we consider the insertion of an intermediate polarizing filter $B$ oriented at $45^\circ$ between filters $A$ ($0^\circ$) and $C$ ($90^\circ$). The light passing through filter $A$ is still polarized at $0^\circ$, with intensity given by Eq.~\eqref{eq:eq01}.

When this light polarized at $0^\circ$ passes through filter $B$, at $45^\circ$, the intensity of the transmitted light is given by Eq.~\eqref{eq:eq05}:
\begin{equation}
\label{eq:eq05}
I_{B} = I_{A} \cos^{2}(45^\circ).
\end{equation}
Since $\cos(45^\circ) = 1/\sqrt{2}$, the transmitted intensity after filter $B$ is given by Eq.~\eqref{eq:eq06}:
\begin{equation}
\label{eq:eq06}
I_{B} = \frac{I_{0}}{2} \times \left(\frac{1}{\sqrt{2}}\right)^{2} = \frac{I_{0}}{4}.
\end{equation}
Now, the light exiting filter $B$ is polarized at $45^\circ$. When passing through filter $C$, oriented at $90^\circ$, the transmitted intensity is again determined by Malus's Law in Eq.~\eqref{eq:eq07}:
\begin{equation}
\label{eq:eq07}
I_{C} = I_{B} \cos^{2}\left(45^\circ\right).
\end{equation}
Substituting $I_{B}$ and $\cos(45^\circ)$, we obtain the final result for the transmitted intensity after filter $C$ (Eq.~\eqref{eq:eq08}):
\begin{equation}
\label{eq:eq08}
I_{C} = \frac{I_{0}}{4} \times \frac{1}{2} = \frac{I_{0}}{8}.
\end{equation}
Therefore, with the intermediate filter at $45^\circ$, the intensity of the light after filter $C$ is not zero, but rather $I_{0}/8$. The intermediate filter allows the light to have a component in the $45^\circ$ direction, which in turn has a component in the $90^\circ$ direction.

\section{Quantum Analysis}

\subsection{Case of Two Perpendicular Filters}

In the formalism of quantum mechanics, the polarization of a photon can be described by vector states in two-dimensional Hilbert space. Consider the filter $A$ at $0^\circ$ corresponds to a horizontal polarization state $\ket{H}$ is represented as:
\begin{equation*}
    \ket{H} = \begin{pmatrix} 1 \\ 0 \end{pmatrix}
\end{equation*}
and, the filter $B$ at $90^\circ$ corresponds to a vertical polarization state $\ket{V}$ is represented as:
\begin{equation*}
    \ket{V} = \begin{pmatrix} 0 \\ 1 \end{pmatrix}.
\end{equation*}

If a photon initially strikes a polarizing filter $A$ aligned at $0^\circ$, the final state after the filter is given by Eq.~\eqref{eq:eq09}:
\begin{equation}
\label{eq:eq09}
\ket{\psi} = \ket{H}.
\end{equation}
The probability of this photon passing through a second filter polarized $B$ at $90^\circ$ is given by the square of the inner product between the initial state $\ket{H}$ and the corresponding state of the filter $\ket{V}$, as indicated in Eq.~\eqref{eq:eq10}:
\begin{equation}
\label{eq:eq10}
P = \left\lvert \braket{V \mid H}\right\rvert^{2}.
\end{equation}
Since $\braket{V \mid H} = 0$, we have $P = 0$. Therefore, the probability of a photon polarized at $A(0^\circ)$ passing through a filter polarized at $B(90^\circ)$ is zero, agreeing with the classical result.

Note, however, that in the classical approach the null value found in Eq.~\eqref{eq:eq03}, refers to the intensity of the light after passing through the polarized filters. In the quantum approach expressed in Eq.~\eqref{eq:eq10}, the null value refers to the probability of some photon passing through the filters. These are different perspectives of the same phenomenon.

\subsection{Case with Intermediate Filter at $45^\circ$}

When we insert an intermediate filter at $45^\circ$, the polarization state after filter $A$ is $\ket{H}$. The filter $B$ at $45^\circ$ corresponds to a diagonal polarization state is a superposition of \(\ket{H}\) and \(\ket{V}\):
\begin{equation*}
    \begin{aligned}
        \ket{D} &= \frac{1}{\sqrt{2}} (\ket{H} + \ket{V}) \\
        &= \frac{1}{\sqrt{2}} \begin{pmatrix} 1 \\ 0 \end{pmatrix} + \frac{1}{\sqrt{2}} \begin{pmatrix} 0 \\ 1 \end{pmatrix} \\
        &= \frac{1}{\sqrt{2}} \begin{pmatrix} 1 \\ 1 \end{pmatrix}
    \end{aligned}
\end{equation*}
The probability of a photon initially in $\ket{H}$ passing through $B$ is given by Eq.~\eqref{eq:eq11}:
\begin{equation}
    \label{eq:eq11}
    \begin{aligned}
        P_{B} &= \left\lvert \braket{D \mid H} \right\rvert^{2} \\
        &= \left\lvert \frac{1}{\sqrt{2}} \braket{H \mid H} + \frac{1}{\sqrt{2}} \braket{V \mid H} \right\rvert^{2} \\
        &= \left\lvert \frac{1}{\sqrt{2}} \right\rvert^{2} = \frac{1}{2}.
    \end{aligned}
\end{equation}
If the photon passes through filter $B$, its state is projected into $\ket{D}$. The probability that this photon passes through filter $C$ at $90^\circ$ (state $\ket{V}$) is given by Eq.~\eqref{eq:eq12}:
\begin{equation}
    \label{eq:eq12}
    \begin{aligned}
        P_{C} &= \left\lvert \braket{V \mid D} \right\rvert^{2} \\
        &= \left\lvert \frac{1}{\sqrt{2}} \braket{V \mid H} + \frac{1}{\sqrt{2}} \braket{V \mid V} \right\rvert^{2} \\
        &= \left\lvert \frac{1}{\sqrt{2}} \times 0 + \frac{1}{\sqrt{2}} \times 1 \right\rvert^{2} = \frac{1}{2}.
    \end{aligned}
\end{equation}
The total probability of transmission through the three filters is the product of the probabilities (Eq.~\eqref{eq:eq13}):
\begin{equation}
\label{eq:eq13}
P_{\text{total}} = P_{B} \times P_{C} = \frac{1}{2} \times \frac{1}{2} = \frac{1}{4}.
\end{equation}
Since the intensity is proportional to the probability, the quantum result is consistent with the classical result of Eq.~\eqref{eq:eq08}.

\section{Discussion \& Conclusion}

The quantum mechanical interpretation of polarization differs from the classical approach primarily in how it describes the state of light and the probabilities associated with its transmission through polarizing filters.

In classical optics, polarization is described using electromagnetic wave theory, where light is treated as a wave that can be polarized in different directions. The intensity of light transmitted through a polarizer is calculated using Malus's Law, which relates the angle of the polarizer to the intensity of transmitted light. The intensity of light after passing through a polarizer is deterministic and can be calculated directly. For instance, when unpolarized light passes through a polarizer, the intensity is halved.

In contrast, quantum mechanics describes the polarization of light in terms of quantum states. For example, the horizontal polarization state, denoted as $\ket{H}$ and the vertical polarization state as $\ket{V}$. The state of a photon can be represented as a vector in a Hilbert space, allowing for a more nuanced understanding of superposition and entanglement. The probability of a photon passing through a polarizer is determined by the inner product of the photon's state vector and the polarizer's state vector.

The inclusion of intermediate filters also highlights the difference. In classical optics, inserting a filter at $45^\circ$ between two perpendicular filters allows some light to pass through, as it can be partially polarized in the direction of the intermediate filter. The transmitted intensity can be calculated using Malus's Law. In quantum mechanics, the intermediate filter alters the state of the photon. The photon can be in a superposition of states after passing through the $45^\circ$ filter, allowing it to have a non-zero probability of passing through the final $90^\circ$ filter. The total probability of transmission through multiple filters is the product of the probabilities for each filter, illustrating the quantum mechanical principles of superposition and projection.

Despite the differences in the description of the classical and quantum analyses, the results of the calculations show an agreement between the two descriptions for the transmission of unpolarized light through polarizing filters. Although at first glance Eqs.~\eqref{eq:eq08} and Eq.~\eqref{eq:eq13} present different numerical results, it should be remembered that the classical result calculates the luminous intensity at the end of each system of polarizing filters, while the quantum version tells of the probability of the photons reaching the end of their path passing through the sequence of filters.

\end{document}